\documentclass[pre,twocolumn,superscriptaddress]{revtex4-1}
\usepackage{amsmath,amssymb}
\usepackage{amsbsy}
\usepackage{graphicx}
\usepackage{subfigure}
\usepackage{color}
\usepackage[plain]{fancyref}
\usepackage{natbib}

\newcommand{\upd}{\mathrm{d}}
\newcommand{\up}[1]{\mathrm{#1}}
\newcommand{\Ei}{\mathrm{Ei}}

\newcommand{\order}{\mathcal{O}}

\renewcommand*{\log}{\ln}
\renewcommand{\bar}{\overline}
\definecolor{mypurple}{RGB}{153,61,113}
\definecolor{myblue}{RGB}{63,61,153}
\definecolor{myokker}{RGB}{153,140,61}
\definecolor{mygreen}{RGB}{61,153,86}
\definecolor{mymarine}{RGB}{61,90,153}
\definecolor{mycyan}{RGB}{0,255,255}

\begin{document}

\title{Persistence of a Brownian particle in a Time Dependent Potential}
\author{D. Chakraborty}
\affiliation{Institute for Theoretical
  Physics, University of Leipzig, \\Vor dem Hospitaltore 1, 04103
  Leipzig, Germany.} 
\altaffiliation[Present address: ]{Theory of Inhomogeneous Condensed
  Matter, Max-Planck-Institute for Intelligent Systems, Stuttgart.}
\thanks{chakraborty@itp.uni-leipzig.de, chakraborty@is.mpg.de.
}

\begin{abstract}
  We investigate the persistence probability of a Brownian particle in
  a harmonic potential, which decays to zero at long times -- leading
  to an unbounded motion of the Brownian particle. We consider two
  functional forms for the decay of the confinement, an exponential
  and an algebraic decay. Analytical calculations and numerical
  simulations show, that for the case of the exponential relaxation,
  the dynamics of Brownian particle at short and long times are
  independent of the parameters of the relaxation. On the contrary,
  for the algebraic decay of the confinement, the dynamics at long
  times is determined by the exponent of the decay. Finally, using the
  two-time correlation function for the position of the Brownian
  particle, we construct the persistence probability for the Brownian
  walker in such a scenario.
\end{abstract}
\maketitle

\section{Introduction}
The phenomenon of persistence has been of continuing interest in the
past decade. Persistence is quantified through the persistence
probability $p(t)$, that a stochastic variable has not changed its
sign over a time $t$. In a wide class of non-equilibrium systems this
probability decays algebraically with an exponent $\theta$ and the
exponent has been studied in systems that include free random walk in
homogeneous \cite{Majumdar1999,Sire2000} and disordered media
\cite{Chakraborty2008}, critical dynamics \cite{Majumdar1996b},
surface growth
\cite{Derrida1995,Krug1997,Kallabis1999,Toroczkai1999,Constantin2004,
  Singha2005,Escudero2009}, polymer dynamics \cite{Bhattacharya2007},
diffusive processes with random initial condition
\cite{Majumdar1996,Derrida1996,Newman1998}, advected diffusive process
\cite{Chakraborty2009} and finance \cite{Zheng2002,Constantin2005}.  A
precise theoretical prediction for $p(t)$ can only be worked out only
for a select few cases \cite{Slepian1962} -- the simplest scenario
being an exponentially decaying stationary correlator, as in the case
of an overdamped Brownian motion. In general, for most Gaussian
stochastic processes the decay of the stationary correlator,
$C(T)\equiv \langle\bar{X}(T)\bar{X}(0)\rangle$, is
non-exponential. The behavior of $C(T)$ in the neighborhood of zero
characterizes the density of zero crossings for the underlying
stochastic process \cite{Majumdar1996, Slepian1962}. When $C(T)$ near
zero, has a quadratic dependence on time in the first order, the
number of zero crossings of the stochastic process is finite, and the
exponent $\theta$ is extracted using the Independent Interval
Approximation (IIA) \cite{Majumdar1996} or the sign time distribution
of the stochastic variable \cite{Newman1998}. Conversely, when
$C(T)\sim 1-\mathcal{O}(T^{\alpha})$, with $\alpha <2$, the density of
zero crossings is infinite and perturbation expansions about a random
walk correlator gives a good estimate of the persistence
exponent\cite{Krug1997}.

The simplest of all these systems, which exhibit an algebraic decay of
$p(t)$ with an exponent $1/2$, is the case of a overdamped Brownian
particle. Lying in the interface of science and engineering, Brownian
motion is ubiquitous around us and plays a dominant role in the nano
and mesoscopic world. The underlying principle of this stochastic
process is not only used for theoretical modeling of a wide range of
complex phenomena \cite{Frey2005a}, but Brownian motion in itself
serves as an experimental tool for probing microscopic environments
\cite{Frey2005a,Wilson2011,Robert2010,Lee2010}.  In the popular Langevin
picture, the erratic motion of a Brownian particle is well described
by Newton's equation of motion with a viscous drag and a
delta-correlated stochastic force acting on the particle.  While the
non-Markovian nature of the phenomenon can be taken into consideration
by using a generalized Langevin equation with a finite correlation
time for the stochastic noise and a memory dependent friction, in
the following discussion we shall restrict ourselves to the Markovian
scenario.

In this article, we investigate the persistence probability of a
Brownian particle in a time dependent potential -- a scenario
corresponding to the trapping of a tracer particle in some potential
which eventually relaxes to zero. To keep the following discussions at
an analytically tractable level, we choose a harmonic potential, given
by $U(x,t)=\frac{1}{2} f(t) x^2$. The function $f(t)$ can be viewed as
time-dependent spring constant, with $f(t)\to 0$ as $t\to \infty$, so
that the particle motion becomes unbounded in the long-time limit. The
converse situation of a constant confinement strength has already been
studied in Ref.~\cite{Slepian1962,Majumdar2001,Chakraborty2007}. In
the Fokker-Planck description, the calculation of the persistence
probability translates to solving the backward Fokker-Planck equation,
with an absorbing wall at the $x=0$. An alternative approach to
determine the survival probability, as outlined in
\cite{Majumdar1999,Majumdar2001,Slepian1962}, is from the two-time
correlation function for the position of the stochastic variable $x$
-- exploiting the fact that for a Gaussian stationary process with a
correlator decaying exponentially at all times, the persistence
probability also decays exponentially.

The rest of the article is organized as follows: we
introduce the dynamical equations of motion and construct the two-time
correlation functions in \fref{sec:msd_discussion}. A discussion on
the mean-square displacement of the Brownian walker and the relevant
time scales due to the time dependent trap is also presented in this
section. We study two types of relaxation phenomena - an exponential
and an algebraic relaxation of the confinement discussed in
\fref{sec:msd_exp_relaxation} and \fref{sec:msd_algb_relaxation},
respectively. Finally, the persistence probability is discussed in
\fref{sec:persistence_probability}.

\section{Brownian particle in a time-dependent potential }
\label{sec:msd_discussion}
For simplicity, we take the overdamped limit for which the
dynamics of a Brownian particle, with a unit mass, is governed by
\begin{equation}
\label{eq:langevin}
\dot{x}=-f(t) x(t) + \eta(t),
\end{equation}
where $\eta(t)$ is the stochastic velocity characterizing the
solvent. The above equation is further supplemented by the 
moments of the stochastic noise,
\begin{equation}
\label{eq:fdt2}
\langle \eta(t) \rangle = 0 \quad \textrm{and} \quad \langle \eta(t) \eta(t')\rangle 
= 2 D \delta(t-t').
\end{equation}
At this point, we assume that the stochastic noise in ``internal'',
characterized by the viscosity and the temperature of the solvent,
while the time-dependent confinement is ``external'' and does not
change the delta correlation. An experimental realization of the model
system would correspond to a laser trapping of a tracer, with the
intensity of laser decaying in time. In such a scenario, the
transport parameters and the temperature $T$
gets renormalized \cite{Rings2010,Chakraborty2011k}. The
corresponding solution to \fref{eq:langevin} is given by
\begin{equation}
\label{eq:x_t}
x(t)=e^{-\int_0^t f(t') \up d \up t'} \int_0^t \up d \up t_1 
\eta(t_1) e^{\int_0^{t_1}  f(t_1') \up d \up t_1'}, 
\end{equation}
with the initial condition $x(0)=0$.
Denoting $g(t)=\int_0^t f(t') \up d \up t'$, the two time correlation
function can be constructed from \fref{eq:x_t},
\begin{eqnarray}
\label{eq:xt1xt2}
\nonumber
\langle x(t_1)x(t_2) \rangle&=e^{-g(t_1)} e^{-g(t_2)}  \int_0^{t_1}
\up d \up t_1'\int_0^{t_2} \up d \up t_2' \\
&\langle \eta(t_1') \eta(t_2')\rangle
e^{g(t_1')} e^{g(t_2')}.
\end{eqnarray}

\subsection{Exponential relaxation}
\label{sec:msd_exp_relaxation}
We first consider the case when the relaxation of the potential is
given by an exponential decay, $f(t)=\lambda e^{-t/\tau}$. There are
two time-scales in the system, $\tau$ and $\lambda^{-1}$. The later
determines the time scale when the Brownian particle is confined in
the potential, whereas the former determines the relaxation of the
potential (\fref{fig:msd}a).  We further consider the
situation when $\tau>\lambda^{-1}$; the relaxation time-scale of the
potential is larger than the entrapment time-scale
$\lambda^{-1}$. When $\tau <\lambda^{-1}$, the confinement decays even
before the particle can be trapped, with the result that the Brownian
particle undergoes free diffusion.

Consequently, the function $g(t)$ takes the form  $g(t)=\lambda \tau
(1-e^{-t/\tau})$. Substituting for $g(t)$ in \fref{eq:xt1xt2} and
subsequently using \fref{eq:fdt2} we arrive at
\begin{eqnarray}
  \label{eq:two_time_corr}
\nonumber
\langle x(t_1) x(t_2) \rangle = 2D  e^{\lambda\tau e^{-t_1/\tau}}
e^{\lambda\tau e^{-t_2/\tau}}\\ 
\int_0^{t_2} \up d \up t_2'\,\, e^{-2\lambda\tau e^{-t_2^{'}/\tau}},
\end{eqnarray}
with the assumption that $t_1>t_2$. Performing the integral over
$t_2'$ in \fref{eq:two_time_corr}, the two-time correlation function
becomes,
\begin{eqnarray}
  \label{eq:two_time_corr_final}
  \nonumber
  \langle x(t_1) x(t_2) \rangle &=&2D \tau\, 
  e^{\lambda\tau e^{-t_1/\tau}}e^{\lambda\tau
    e^{-t_2/\tau}} \bigr[ \Ei(-2 \lambda \tau) \\
\nonumber
  &-& \Ei(-2\lambda\tau e^{-t_2/\tau}) \bigr],\\
\end{eqnarray}
where $\Ei(x)$ is the exponential integral defined as 
\begin{equation}
  \label{eq:ei_fn}
  \Ei(t)=-\int_{-t}^\infty z^{-1}e^{-z} \; \upd z.
\end{equation}

Using \fref{eq:two_time_corr_final}, the mean-square displacement
$\langle x^2(t) \rangle$ reads as,
\begin{equation}
  \label{eq:msd_exp_decay}
  \langle x^2(t) \rangle =2D\tau \,e^{2 \lambda \tau e^{-t/\tau}} \bigr[
  \Ei(-2\lambda\tau)
-\Ei(-2\lambda\tau e^{-t/\tau})\bigr].
\end{equation}
At this point, it is instructive to construct the limiting behaviors
of the mean-square displacement -- when $t<\lambda^{-1}$ and $t>\tau$.
There are two scenarios we consider below -- the first, when $\tau$ is
large so that the limit $\tau \to \infty$ is appropriate, and the
second when $\tau$ remains finite. In the limit of $\tau \to \infty$,
the relaxation of the potential is slow and the Brownian particle
feels a constant confinement strength $\lambda$ and
\fref{eq:msd_exp_decay} reduces to
\begin{equation}
  \label{eq:tau_infty}
  \langle x^2(t) \rangle= \frac{D}{\lambda}(1-e^{-2t\lambda}) + 
\order(\tau^{-1}).
\end{equation}
To construct the corresponding two-time correlation function, we
expand the exponentials in \fref{eq:two_time_corr} and keep
the terms which are independent of $\tau$. The evaluation of the
integral over $t_2'$ then gives,
\begin{equation}
  \label{eq:two_time_tau_infty}
  \langle x(t_1) x(t_2) \rangle
  =\frac{D}{\lambda}\left[e^{-\lambda(t_1-t_2)}-e^{-\lambda (t_1+t_2)}\right],
\end{equation}
which is exactly the correlation function for a non-stationary
Ornstein-Uhlenbeck process \cite{Chakraborty2007}.  Eventually, for
$t>>\tau$ the particle motion becomes unbounded and the mean-square
displacement grows linearly with time.  A formal quantitative result
in this limit can be derived if we take the limit of $t \to \infty$ in
\fref{eq:msd_exp_decay} and expand the term within the brackets to get
\begin{eqnarray}
  \label{eq:long_time_limit}
  \nonumber
  &\langle x^2(t)\rangle& =2D\tau \, e^{2\lambda \tau e^{-t/\tau}}\bigr[ -\gamma +
  2\lambda \tau e^{-t/\tau} - \lambda^2 \tau^2 e^{-2 t/\tau}\\
  \nonumber
  &+&\Ei(-2\lambda \tau) + \frac{t}{\tau} + 
 \frac{1}{2} \log(1/2\lambda^2 \tau) \bigr],\\
\end{eqnarray}
where $\gamma$ is the Euler's constant with a numerical value of $\sim
0.5772$. Keeping in mind that $t>>\tau$, the exponential functions in
\fref{eq:long_time_limit} can be ignored in comparison to the linearly
growing term which survives, so that we recover the classic diffusion
of the Brownian particle with $\langle x^2(t) \rangle = 2D t$. In the
opposite limit of $t \to 0$, a Taylor expansion of
\fref{eq:msd_exp_decay} yields,
\begin{equation}
  \label{eq:short_time_limit}
  \langle x^2(t) \rangle = 2D t+\order(t^2).
\end{equation}
The two limiting behaviors in \fref{eq:long_time_limit} and
\fref{eq:short_time_limit} are completely independent of the time
scales, and therefore, do not contain any information about the
confinement potential. On the contrary, when the relaxation is slow
($\tau \to \infty$), only the short-time dynamics is independent of
$\lambda$ or $\tau$. The asymptotic mean-square displacement, in the
limit of a slow relaxation, is constant in time and is determined by the
ratio $D/\lambda$.

\subsection{Algebraic relaxation}
\label{sec:msd_algb_relaxation}
We now consider our second choice for the relaxation dynamics of the
harmonic potential -- an algebraic decay of the time dependent spring
constant,
\begin{equation}
  \label{eq:conf_pot_algb}
  U(x,t)=\frac{\lambda}{2}\left(\frac{\tau}{t}\right)^{\alpha} x^2,
\end{equation}
with $\alpha \leq 1$. Using \fref{eq:xt1xt2}, the two-time correlation
function is
\begin{eqnarray}
  \label{eq:two_time_algb}
\nonumber
  \langle x(t_1) x(t_2) \rangle = \frac{2 D}{1-\alpha}\, e^{-\lambda_1
    t_1^{1-\alpha}} e^{-\lambda_1 t_2^{1-\alpha}}
  \int_0^{t_2^{1-\alpha}}\up d \up u \\
u^{\alpha/(1-\alpha)} e^{2\lambda_1 u} \,\,\, \,
\end{eqnarray}
where $\lambda_1=\lambda \tau^{\alpha}/(1-\alpha)$. The integral
\fref{eq:two_time_algb} over $\up u$ yields 
\begin{eqnarray}
  \label{eq:two_time_algb_final}
  \nonumber
  \langle x(t_1) x(t_2) \rangle =  \frac{2 D}{1-\alpha}\, e^{-\lambda_1
    t_1^{1-\alpha}} e^{-\lambda_1 t_2^{1-\alpha}}\\
  \nonumber
 (-2 \lambda_1)^{-1/(1-\alpha)}\;  \gamma
 \left(\frac{1}{1-\alpha}, -2\lambda_1 t_2^{1-\alpha} \right)
\\
 \end{eqnarray}
where the integral $\gamma$ is the lower incomplete gamma function defined as
\begin{equation}
  \label{eq:en_function}
  \gamma(a,z)=\int_0^{z} e^{-u} u^{a-1} \up d \up u
\end{equation}
and $\Gamma(a,z)$ is the upper incomplete Gamma function satisfying
$\Gamma(a,z)+\gamma(a,z)=\Gamma(a)$. The numerical value of
$\gamma(a,z)$ can be evaluated using Gauss's continued fraction, which
converges for all values of $z$. 
\begin{figure*}
\includegraphics[width=0.45\linewidth]{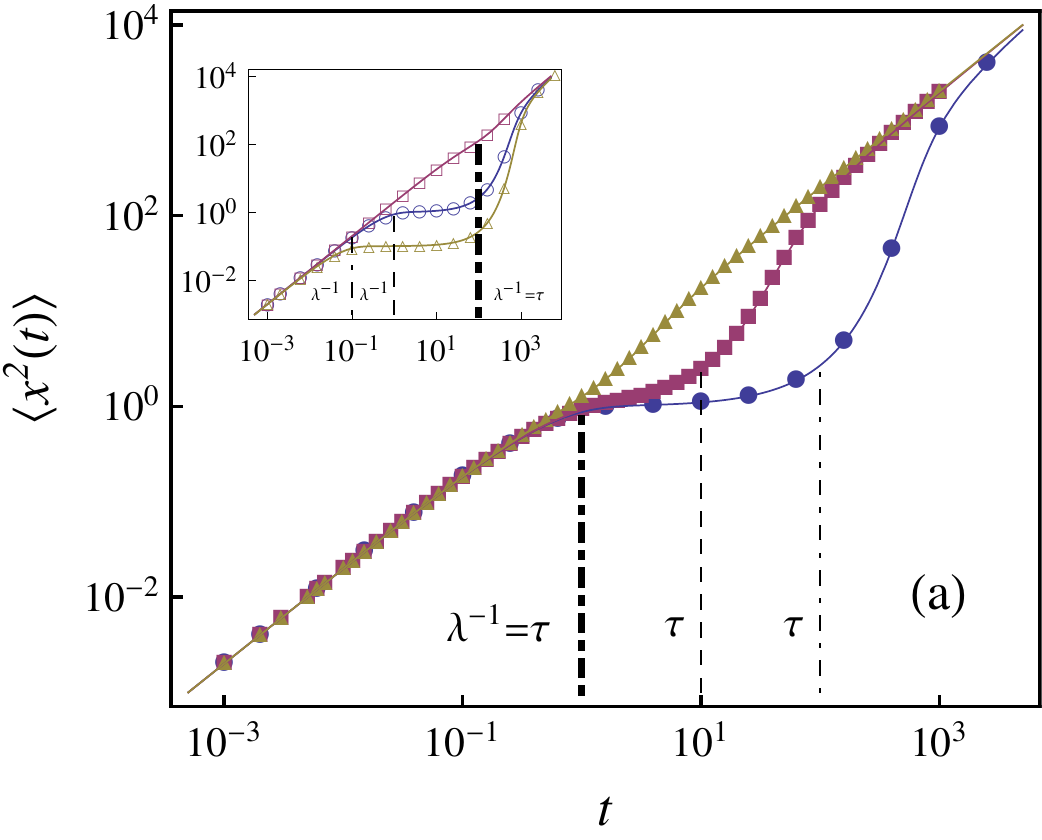}
\hfill
\includegraphics[width=0.45\linewidth]{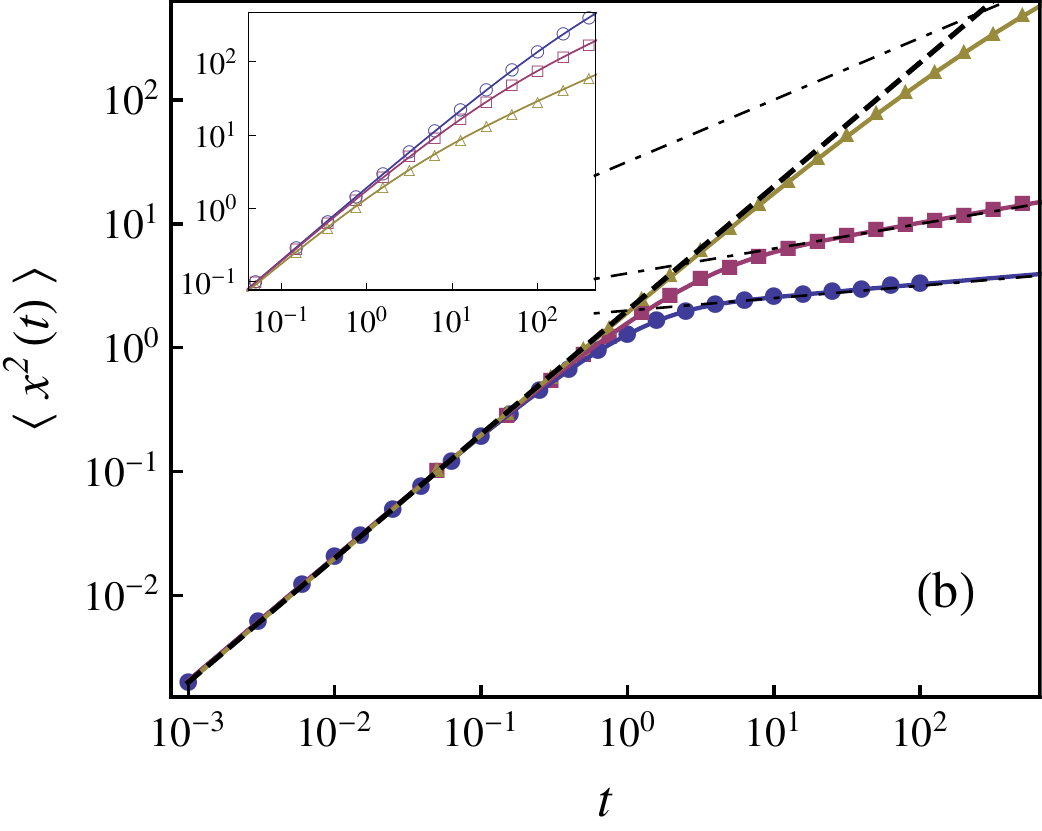}
\hfill
\caption{(Color online) \textbf{(a):} Double logarithmic plot of the
  mean-square displacement $\langle x^2 (t) \rangle$ for the
  exponential relaxation of the confinement; with $\lambda=1$ and
  $\tau=1.0$ ({\color{myokker} $\blacktriangle$}), $10.0$
  ({\color{mypurple} {\footnotesize $\blacksquare$}}) and $100.0$
  ({\color{myblue} {\Large $\bullet$}}). The solid lines are the plots
  of \fref{eq:msd_exp_decay} for the corresponding values of $\tau$
  and $\lambda$. The thick dot-dashed line in the main plot
  corresponds to $\lambda^{-1}=\tau=1$, while the dashed and the
  dot-dashed line corresponds to $\tau=10$ and $\tau=100$,
  respectively. \emph{\textbf{Inset:}} We depict the variation of the
  mean-square displacement for different values of $\lambda$ with
  $\tau=100$ fixed; $\lambda = 0.01$ ({\color{mypurple} {\footnotesize
      $\square$}}), $1.0$ ({\color{myblue} {\Large $\circ$}}) and
  $10.0$ ({\color{myokker} {$\vartriangle$}}). The solid lines are
  plot of \fref{eq:msd_exp_decay} for the corresponding values of
  $\lambda$ and $\tau$. The thick dot-dashed line corresponds to
  $\lambda^{-1}=\tau=100$ while the dashed lines denote the values of
  $\lambda^{-1}=0.1$ and $\lambda^{-1}=1$.  \textbf{(b):} Plot of
  mean-square displacement for $\lambda=1.0, \tau=0.001$ and
  $\alpha=0.1$ ({\color{myblue} {\large $\bullet$}}), $0.2$
  ({\color{mypurple} {\footnotesize $\blacksquare$}}) and $0.5$
  ({\color{myokker} {$\blacktriangle$}}) and the corresponding plots
  of \fref{eq:msd_algb_decay} for the three values of $\alpha$. The
  thick dashed black line is the plot of $2Dt$ and the thin dot-dashed
  black lines are the plots of \fref{eq:msd_algb_long_time} for the
  corresponding values of $\alpha$. \emph{\textbf{Inset:}} Plot of
  mean-square displacement for $\lambda=1.0, \alpha=0.5$ but with
  $\tau=0.001$ ({\color{myblue} {\large $\circ$}}), $0.01$
  ({\color{mypurple} {\footnotesize $\square$}}) and $0.1$
  ({\color{myokker} {$\vartriangle$}}).  }
\label{fig:msd}
\end{figure*}

Substituting $t_1=t_2=t$, the mean-square displacement is given by
\begin{eqnarray}
  \label{eq:msd_algb_decay}
\nonumber
  \langle x^2(t) \rangle = \frac{2D}{1-\alpha}\,e^{-2\lambda_1
  t^{1-\alpha}} (-2 \lambda_1 )^{-1/(1-\alpha)} \;\\
\nonumber
 \gamma
 \left(\frac{1}{1-\alpha}, -2\lambda_1 t_2^{1-\alpha} \right)\\
\end{eqnarray}

Unlike the exponential relaxation of the potential, there is a single
crossover time scale which emerges from \fref{eq:msd_algb_decay} 
\begin{equation}
  \label{eq:crossover_time}
  \bar{\tau}=\left(\frac{1-\alpha}{\lambda \tau^{\alpha}}\right)^{1/(1-\alpha)},
\end{equation}
and separates the regimes of normal diffusion and sub-diffusion in the
system (\fref{fig:msd}b ). A Taylor expansion of
\fref{eq:msd_algb_decay} for $t < \bar{\tau}$ gives,
\begin{equation}
  \label{eq:msd_algb_short_time_limit}
  \langle x^2(t) \rangle = 2Dt+ \order (t^{2-\alpha}) \quad \mathrm{for}
  \quad t<\bar{\tau},
\end{equation}
while the asymptotic expansion 
yields
\begin{equation}
  \label{eq:msd_algb_long_time}
  \langle x^2(t) \rangle = \left(\frac{D}{\lambda \tau^\alpha}\right)
  t^\alpha +\order(t^{-(1-2\alpha)}) \quad \mathrm{for} \quad t>\bar{\tau}. 
\end{equation}
This counter intuitive result can be understood by considering the
motion of a free Brownian particle. In the absence of the confinement
potential the Brownian particle moves a distance $\sqrt{t}$ in time
$t$. If we now switch on the potential, the strength of the potential
becomes $\lambda (\tau/t)^{\alpha} t$ and for $\alpha<1$ we see that
the particle feels the ``soft'' walls all the time. Mathematically,
this argument translates to the fact that the new time scale
$\bar{\tau}$ in \fref{eq:crossover_time} diverges as $\alpha \to 1$
and is not defined in the real line for $\alpha >1$. We note
that for $\alpha=0$, eq.\fref{eq:msd_algb_decay} reduces to that of the
Ornstein-Uhlenbeck process,
\begin{equation}
  \label{eq:msd_algb_decay_alpha0}
  \langle x^2(t) \rangle= \frac{D}{\lambda}\bigr[ 1- e^{-2 \lambda t}\bigr].
\end{equation}

In \fref{fig:msd}, we show the mean-square displacement of a Brownian
particle whose dynamics is governed by \fref{eq:langevin} and
\fref{eq:fdt2}, with $f(t)$ given by an exponential (\fref{fig:msd}a)
and an algebraic (\fref{fig:msd}b) relaxation. The
numerical integration of \fref{eq:langevin} was done using the Euler
scheme with an integration time step of $\up d \up t=0.001$. In the
numerical solutions, the value of the diffusion coefficient $D$ was
taken as unity.  For the exponential relaxation of the confinement,
the measured mean-square displacement show three distinct regimes --
two diffusive regimes with a crossover in between. For very short
($t<\lambda^{-1}$) and long times ($t>\tau$), the particle does not
feel the trap and its motion is purely diffusive, corresponding to
\fref{eq:long_time_limit} --\fref{eq:short_time_limit} . In the
intermediate times, we observe a plateau for $ \lambda^{-1}<t<\tau$,
corresponding to the trapping of the particle in the potential. To
understand the origin of this plateau, we expand the exponential in
\fref{eq:langevin}, and retaining the zeroth order term then leads to
a constant confinement, so that the mean-square displacement saturates
to a value $\propto \lambda^{-1}$. This behavior can be observed in
the left panel of \fref{fig:msd}, the main figure of which presents
data for constant $\lambda$ but different $\tau$, while the inset
shows data for a constant $\tau$ but different $\lambda$. A comparison
shows that the plateau is determined by $\lambda^{-1}$. On the
contrary, for the algebraic relaxation, since a single time scale
emerges from the dynamics, we observe only one crossover regime
determined by $\bar{\tau}$ which separates the diffusive and the
sub-diffusive regimes (the right panel of \fref{fig:msd}).

\section{Persistence probability}
\label{sec:persistence_probability}
To obtain the persistence probability, we take the route outlined in
Ref.~\cite{Majumdar1999,Slepian1962} -- we map the non-stationary
process $x(t)$ to a stationary Ornstein-Uhlenbeck process.  This is
usually achieved, first by a normalization of $x(t)$ by $\sqrt{\langle
  x^2(t)\rangle}$, the root-mean-square distance the particle has
traveled and then using a suitable transformation in time. Once, we
have the stationary process $\bar{X}$, with correlator $C(T)$, the
persistence problem reduces to calculation of no zero crossing of
$\bar{X}$. When $C(T)$ is a purely exponential decay for all times,
the persistence probability is the solution to the backward
Fokker-Planck equation for an Ornstein-Uhlenbeck process, which can be
shown to decay as $P(T)=\frac{2}{\pi}\sin^{-1}\left[C(T)\right]$,
\cite{Majumdar1999, Majumdar2001}. An application of this method,
therefore, requires the transformation of the stochastic process
$x(t)$ to a Gaussian stationary process. Since the correlation
function in \fref{eq:two_time_corr_final} and
\fref{eq:two_time_algb_final} is non-stationary, we make the following
transformations -- we first define the normalized variable $\bar{X}(t)
\equiv \frac{x(t)}{\sqrt{\langle x^2(t) \rangle}}$ and construct the
two-time correlation function, following which we make a suitable
transformation in time to make the correlator stationary, as well as
an exponentially decaying function for all times.

Before we proceed to give a derivation of the persistence probability
for the two models introduced above, we derive a general result
applicable to the model system in \fref{eq:langevin}. To transform the
nonstationary process in \fref{eq:langevin}, we consider the
transformations $\bar{X}=x(t)/l(t)$ and $e^T=l^2(t)e^{2g(t)}$, where
$l^2(t)=\langle x^2(t) \rangle$ and $g(t)=\int f(t') \upd
t'$. Substituting these transformations in \fref{eq:langevin}, 
we obtain a stationary Ornstein-Uhlenbeck process,
\begin{equation}
  \label{eq:transformed_ou}
  \frac{\upd \bar{X}}{\upd T}=-\frac{1}{2}\bar{X}+\bar{\eta}(T),
\end{equation}
where $\bar{\eta}(T)$ is a Gaussian white noise with zero mean and
unit variance. The relation between $\bar{\eta}$ and $\eta$ can be
determined from the transformation of the delta function and takes the
form $\bar{\eta}(T)=\frac{l(t)}{2D_0}\eta(t)$.  The stationary
correlator for the process in \fref{eq:transformed_ou} is then given
by $C(T)=e^{-T/2}$. Accordingly, the persistence probability in real
time decays as $p(t) \sim e^{-g(t)}/l(t)$. In the following, we illustrate this
explicitly for the two specific cases presented in
\fref{sec:msd_exp_relaxation} and \fref{sec:msd_algb_relaxation}.

\subsection{Exponential Relaxation}
The two-time correlation function for $\bar{X}$ reads,
\begin{equation}
  \label{eq:corr_norm_var_def}
  \langle \bar{X}(t_1) \bar{X}(t_2) \rangle = \dfrac{\langle x(t_1)
    x(t_2) \rangle}{\sqrt{\langle x^2(t_1)\rangle \langle x^2(t_2)\rangle }}
\end{equation}
Using \fref{eq:two_time_corr_final} and \fref{eq:msd_exp_decay}
in the above equation, we have
\begin{equation}
  \label{eq:two_time_norm_var}
  \langle \bar{X}(t_1) \bar{X}(t_2) \rangle = \sqrt{\dfrac{\Ei(-2
      \lambda \tau )-\Ei(-2\lambda \tau e^{-t_2/\tau}) }{\Ei(-2
      \lambda \tau )-\Ei(-2\lambda \tau e^{-t_1/\tau})}}.
\end{equation}

The time transformation $e^T=l^2(t)e^{2g(t)}$ reads,
\begin{equation}
  \label{eq:time_transformation_exp_decay}
  e^{T}=\Ei(-2 \lambda \tau ) -\Ei(-2 \lambda \tau e^{-t/\tau}),
\end{equation}
which transforms \fref{eq:two_time_norm_var} to 
\begin{equation}
  \label{eq:corr_norm_var}
  \langle \bar{X}(T_1) \bar{X}(T_2) \rangle= e^{-\frac{1}{2}(T_1 -T_2)}
\end{equation}

\begin{figure}
\includegraphics[width=\linewidth]{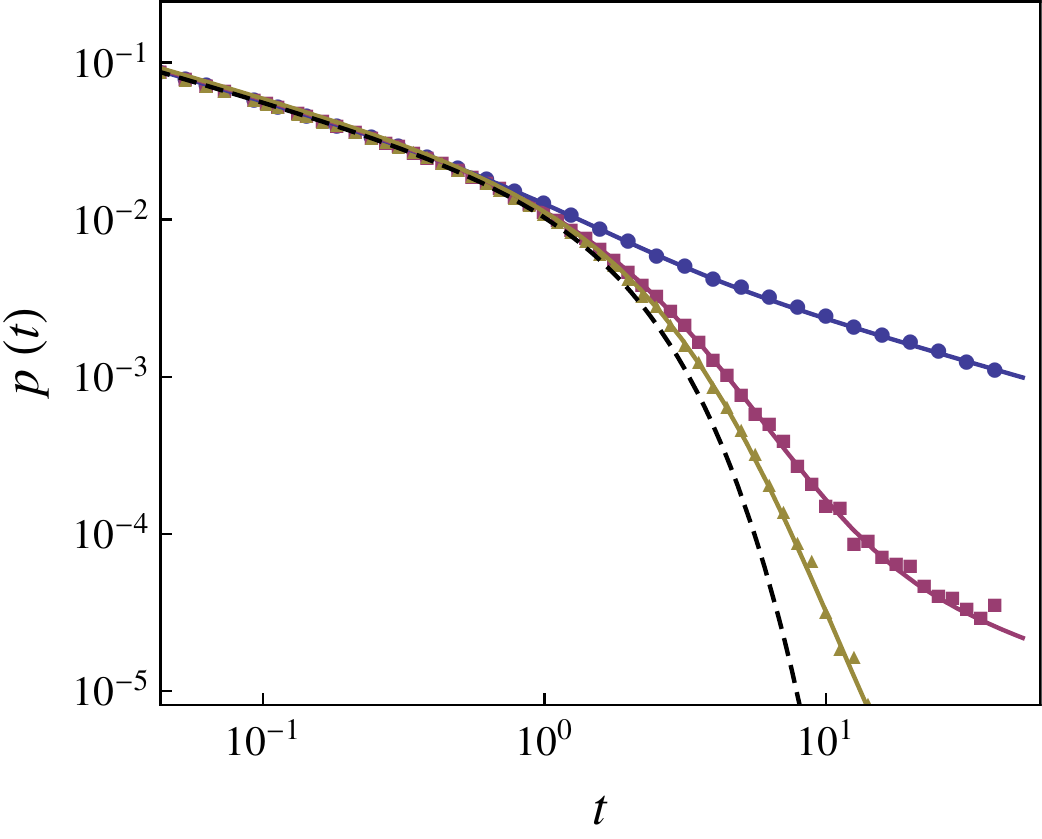}
\caption{(Color online) Double logarithmic plot of persistence
  probability where $f(t)$ decays exponentially, for $\lambda=1.0$ and
  $\tau=1.0$ ({\color{myblue} {\large $\bullet$}}), $5.0$
  ({\color{mypurple} {\footnotesize $\blacksquare$}}) and $10.0$
  ({\color{myokker} {$\blacktriangle$}}). The solid and the dashed
  lines are the plots of \fref{eq:per_exp_decay} for the corresponding
  values of $\tau$ and dashed line is the plot of
  \fref{eq:per_prb_tau_infty}.}
\label{fig:perprob_exp_decay}
\end{figure}

The correlator for the stochastic process $\bar{X}$ is stationary and
exponentially decaying. The asymptotic behavior of the persistence
probability for such a process is then given by $P(T) \sim e^{-T/2}$
\cite{Slepian1962}. Transforming back to real time, the persistence
probability for the process $x(t)$ is then given by,
\begin{equation}
  \label{eq:per_exp_decay}
  p(t)\sim \left[\Ei(-2\lambda \tau)-\Ei(-2\lambda 
    \tau e^{-t/\tau})\right]^{-1/2}
\end{equation}
For $t<<\lambda$ and $t>>\tau$, a Taylor and an asymptotic expansion
of the above equation gives $p(t) \sim t^{-1/2}$. Finally, in the
limit of $\tau \to \infty$, the persistence probability reads as
\begin{equation}
  \label{eq:per_prb_tau_infty}
  p(t) \sim \left[(1-e^{-2 \lambda t})e^{2 \lambda t} 
    +\order(\tau^{-2})\right]^{-1/2}
\end{equation}
which is identical to the result of Ref.\cite{Chakraborty2007}. To
determine the persistence probability of the Brownian particle using a
numerical integration, we chose an ensemble of random initial
conditions in the neighborhood of zero (so that the sign of $x(0)$ is
well defined) and followed the sign change of
the position. The fraction of particles which did not change the sign
of the coordinates in time $t$ gives an estimate of the persistence
probability.  The results presented in \fref{eq:per_exp_decay} (the
colored lines in \fref{fig:perprob_exp_decay}) and
\fref{eq:per_prb_tau_infty} (the black dashed line in
\fref{fig:perprob_exp_decay}) is compared with the measured
persistence probability using the numerical simulation of
\fref{eq:langevin} (the solid points) in
\fref{fig:perprob_exp_decay}. For short and long times, the
persistence probability $p(t) \sim t^{-1/2}$, a signature of purely
diffusive motion presented in \fref{eq:long_time_limit} and
\fref{eq:short_time_limit}.

\subsection{Algebraic relaxation}

To determine the survival probability, we proceed in the similar way
and construct the two-time correlation function for the normalized
variable $\bar{X}$.
\begin{equation}
  \label{eq:two_time_corr_norm_var_algb}
    \langle \bar{X}(t_1) \bar{X}(t_2)\rangle= \sqrt{\frac{h(t_2)}{h(t_1)}}
\end{equation}
where the function $h(t)$ is the bracketed term in \fref{eq:msd_algb_decay},
\begin{eqnarray}
  \label{eq:function_g}
\nonumber
 h(t)=(-2\lambda_1)^{1/(1-\alpha)}\;\;
 \gamma\left(1/(1-\alpha),-2\lambda_1 t^{1-\alpha} \right)
\end{eqnarray}
\begin{figure}
\includegraphics[width=\linewidth]{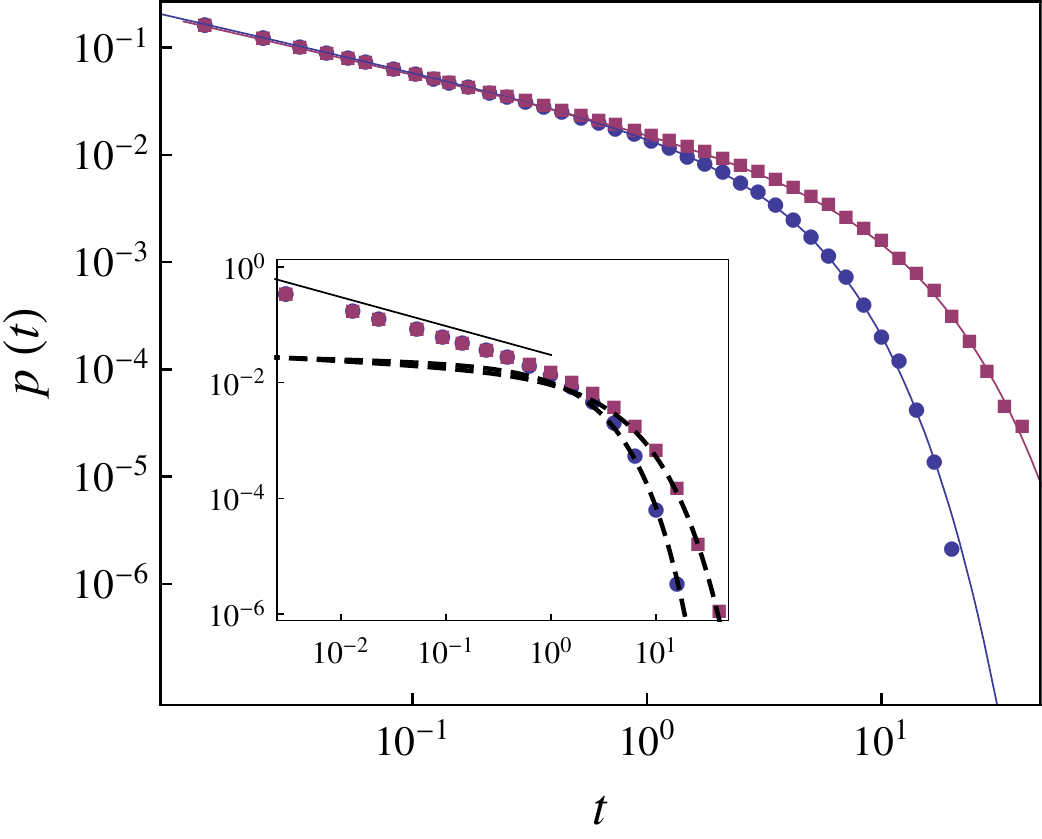}
\caption{(Color online) Plot of persistence probability for
  $\lambda=1.0, \tau=0.001$ $\alpha=0.1$ ({\color{myblue} {\Large
      $\bullet$}}) and $0.2$ ({\color{mypurple} {\footnotesize
      $\blacksquare$}}). The solid lines in the main plot are 
  plots of \fref{eq:per_prob_algb_decay} for the corresponding values
  of $\alpha$. \emph{\textbf{Inset:}} Plot of the persistence
  probability for $\lambda=1, \tau=0.01$ with $\alpha=0.1$
  ({\color{myblue} {\Large $\bullet$}}) and $0.2$ ({\color{mypurple}
    {\footnotesize $\blacksquare$}}). The solid black line is the plot
  of $t^{-1/2}$ and the dashed lines are the plots of the asymptotic
  expansion \fref{eq:per_prob_long_time_algb_decay}. }
\label{fig:perprob_algb_decay}
\end{figure}

Defining the time transformation $e^T \equiv l^2(t)e^{2g(t)}= h(t)$, the
non-stationary correlator in \fref{eq:two_time_algb_final} is
transformed into a Gaussian stationary correlator which decays
exponentially. Following \cite{Slepian1962}, the persistence
probability, in real time decays as
\begin{equation}
  \label{eq:per_prob_algb_decay}
\begin{split}
  p(t) \sim \biggr[ (-2\lambda_1)^{-1/(1-\alpha)} \;\;
 \gamma\left(1/(1-\alpha),-2\lambda_1 t^{1-\alpha} \right) \biggr]^{-1/2}
\end{split}
\end{equation}

We next consider the limiting behaviors of the persistence probability
given in \fref{eq:per_prob_algb_decay}. Substituting $\alpha=0$ in
\fref{eq:per_prob_algb_decay}, the probability reduces to
the case of a harmonically confined Brownian particle with constant
confinement strength \cite{Chakraborty2007},
\begin{equation}
  \label{eq:per_prob_alpha0}
  p(t) \sim \left[ e^{2 \lambda t} (1-e^{-2 \lambda t})\right]^{-1/2}
\end{equation}
For a finite value of $\alpha <1$, when $t<\bar{\tau}$, the persistence
probability decays as $p(t) \sim t^{-1/2}$ while an asymptotic
expansion of \fref{eq:per_prob_algb_decay} gives,
\begin{equation}
  \label{eq:per_prob_long_time_algb_decay}
p(t) \sim \dfrac{1}{t^{\alpha/2}}e^{-{(t/\tau)}^{1-\alpha}}.
\end{equation}
In \fref{fig:perprob_algb_decay}, we compare the results
of\fref{eq:per_prob_algb_decay},\fref{eq:per_prob_alpha0} and
\fref{eq:per_prob_long_time_algb_decay} with the measured persistence
probability from the numerical integration of \fref{eq:langevin}. The
colored lines in the figure correspond to
Eq.\fref{eq:per_prob_algb_decay}, while the dot-dashed lines are 
plots of Eq.\fref{eq:per_prob_long_time_algb_decay}. At short times,
the motion is purely diffusive, and therefore we observe a $t^{-1/2}$
decay of $p(t)$ (the solid line in \fref{fig:perprob_algb_decay}).

We note, that even though the mean-square displacement for $t >> \tau$
is similar to that of fractional Brownian motion, the decay of the
persistence probabilities in the two scenarios are entirely
different. For a particle which performs a fractional Brownian motion,
the corresponding steady state persistence probability decays purely
algebraically with an exponent $1-\alpha/2$ \cite{Krug1997}. 

\subsection{Effect of Inertia}

Finally, before concluding, we remark upon the divergence of $p(t)$ as
$t \to 0$. This singularity is entirely the artifact of
coarse-graining in \fref{eq:langevin}, where we have neglected the
inertia term. Strictly speaking, at this level of coarse-graining, we
are not allowed to take the $t \to 0$ limit, since the inertia of the
particle plays a important role at such short times. The inclusion of
the inertia term changes the short-time dynamics of the particle at to
a deterministic one, as opposed to purely diffusive motion observed in
the overdamped limit.  Since the motion is now deterministic,
the particle is persistently driven away from its initial position
(the velocities remain strongly correlated), with the effect that the
survival probability becomes constant. The purpose of this section is
not only to demonstrate this fact that the inertia term in the
Langevin equation indeed removes the singularity in the persistence
probability, but is also motivated by the recent experimental evidence
of the ballistic regime of a Brownian particle
\cite{Huang2011,Li2010}. While an accurate analysis would correspond
to solving \fref{eq:langevin} with the inertia term included, but it
becomes difficult to extract any information from the resulting
expressions.  However, since we are looking at a time much shorter
than $\lambda^{-1}$, exclusion of the confinement is justified -- as
the particle does not feel the confinement at such small times.

To this end, we consider the complete Langevin equation for the
momentum of a particle without any potential confinement,
\begin{equation}
  \label{eq:full_langevin}
  \dot{p}=-\frac{\gamma}{m} p +\eta 
\end{equation}
together with $\; \langle \eta(t) \rangle=0$ and $\langle \eta(t)
\eta(t') \rangle=2k_B T \gamma \delta(t-t')$. Since at short times,
the dynamics of a Brownian particle in the model systems presented
above is purely diffusive, it suffices to consider the Langevin
equation for a free particle for our present discussion.
\begin{figure}[!t]
\includegraphics[width=\linewidth]{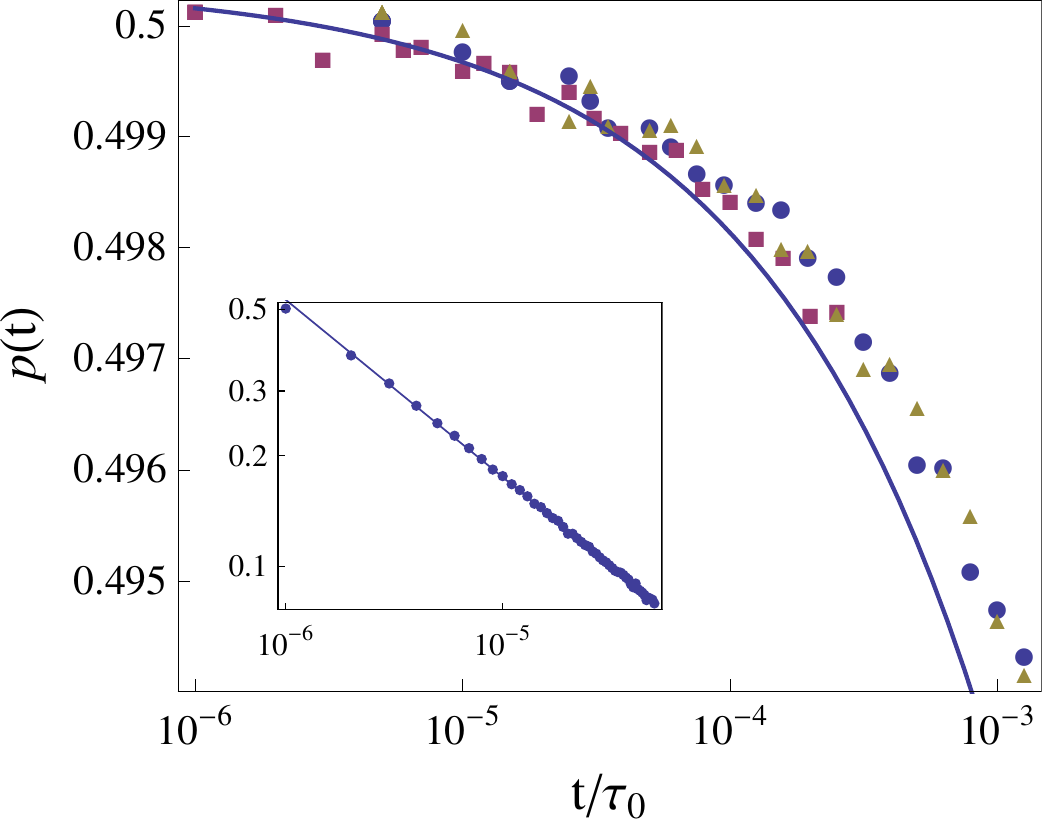}
\caption{(Color online) The main figure depicts a double logarithmic plot of
  persistence probability of a Brownian walker with the inertia term
  included in the Langevin equation for the ratio of $\gamma/m=0.2$
  ({\color{mypurple}$\blacksquare$}), $1.0$
  ({\color{myblue} {\Large $\bullet$}}) and $5.0$
  ({\color{myokker}$\blacktriangle$}). The solid line is a plot of
  the approximate result derived in
  \fref{eq:per_prb_langevin_full}. \emph{\textbf{Inset:}} Double
  logarithmic plot of the persistence probability of a Brownian
  particle in the overdamped limit. The solid line is power law fit to
  the data $t^{-\theta}$, with $\theta=0.482422$. The analytical
  prediction for $\theta$ in this limit is $1/2$.}
\label{fig:perprob_free_bm}
\end{figure}

The two-time correlation function for the velocities decays
exponentially as $\langle v(t_1)v(t_2)\rangle =\frac{k_B T}{m} e^{-
  |t_1 - t_2|/\tau_0}$ (with an initial condition $\langle
v^2(0)\rangle=\frac{k_B T}{m}$) and the position correlation,
\begin{eqnarray}
  \label{eq:pos_corr_langevin}
\nonumber
\langle x(t_1) x(t_2) \rangle &=& \frac{k_B T}{m} \biggr[ \frac{2 m}{\gamma}t_2-
\frac{m^2}{\gamma^2} \\
\nonumber
&+&\frac{m^2}{\gamma^2}\left(e^{-t_1/\tau_0}
+e^{- t_2/\tau_0}-e^{-(t_1 -t_2)/\tau_0}\right)\biggr] \\
\end{eqnarray}
with $\tau_0=m/\gamma$ and the assumption that $t_1>t_2$. While the
transformation of the correlator in \fref{eq:pos_corr_langevin} to a
stationary process is nontrivial, we can still extract some
information about the persistence probability in the limit of $t\to
0$. Keeping in mind this limit and that $t_1>t_2$, a Taylor expansion
of \fref{eq:pos_corr_langevin} yields,
\begin{equation}
  \label{eq:taylor_expansion_xt1xt2}
  \langle x(t_1) x(t_2) \rangle = \frac{t_1 t_2}{m} \left(1-\frac{1}{2}
    \frac{t_1}{\tau_0}\right)
\end{equation}
The two-time correlation function $\langle \bar{X}(t_1) \bar{X}(t_2)
\rangle$ reads as,
\begin{equation}
  \label{eq:two_time_langevin_full}
  \langle \bar{X}(t_1) \bar{X}(t_2) =\sqrt{\frac{1-
      t_1/2\tau_0}{1-t_2/2\tau_0}}
\end{equation}
The transformation $e^T=(1- t/2\tau_0)^{-1}$ transforms
\fref{eq:two_time_langevin_full} into a stationary process with an
exponentially decaying correlation function. The persistence
probability $p(t)$ in real time the translates to,
\begin{equation}
  \label{eq:per_prb_langevin_full}
  p(t) \sim \frac{2}{\pi} \sin^{-1}[\sqrt{1- t/2\tau_0}]
\end{equation}

The numerical integration of \fref{eq:full_langevin} was done with an
implicit integration scheme based on the Leap-Frog algorithm for
different values of the ratio $\gamma/m$ and the value of $T$ was
set to unity. The results of the simulations are presented in
\fref{fig:perprob_free_bm} and compared to the approximate formula for
$p(t)$ in \fref{eq:per_prb_langevin_full}.

\section{Conclusion}
In conclusion, we have investigated the persistence probability of a
harmonically confined Brownian particle in the overdamped limit, with
the potential relaxing to zero at long times.  We consider two
functional forms of the relaxation -- an exponential and an algebraic
relaxation. The simple model system presented in this article is
analogous to a moving wall \cite{Redner2001}, with a ``hard'' wall
replaced by a ``soft'' wall. The external confinement can be realized
using a laser-trapping experiment, with the intensity of the laser
decaying in time. When the confining potential relaxes exponentially,
we observe that the dynamics of the Brownian particle at short and
long times is purely diffusive and independent of the relaxation time
scales. On the other hand, for an algebraic relaxation, the motion at
long times is determined by the exponent of the relaxation. Using the
two-time correlation function for the position of the Brownian
particle, we construct the persistence probability of the Brownian
particle in the two scenarios.

\section{Acknowledgment}
This work was funded by the Alexander von Humboldt foundation. The
author gratefully acknowledges fruitful discussions with Klaus Kroy
(University of Leipzig), Jakob Bullerjahn (University of Leipzig),
Sebastian Sturm (University of Leipzig) and Debasish Chaudhuri (IIT
Hyderabad) and thank them for a careful reading of the manuscript.
\bibliography{library_per}
\bibliographystyle{apsrev4-1}
\end{document}